\documentclass[aps,amsmath,amssymb,10pt,showpacs]{revtex4}
\usepackage{graphicx}
\usepackage{mathrsfs}

\begin{document}

\title{Unitarity and the Hilbert space of quantum gravity}
\author{Stephen~D.~H.~Hsu} \email{hsu@uoregon.edu}\affiliation{Institute of Theoretical Science, University of Oregon,
Eugene, OR 97403 }
\author{David Reeb\,} \email{dreeb@uoregon.edu}\affiliation{Institute of Theoretical Science, University of Oregon,
Eugene, OR 97403 }

\date{September 2008}

\begin{abstract}
Under the premises that physics is unitary and black hole evaporation is complete (no remnants, no topology change), there must exist a one-to-one correspondence between states on future null and timelike infinity and on any earlier spacelike Cauchy surface (e.g., slices preceding the formation of the hole). We show that these requirements exclude a large set of semiclassical spacetime configurations from the Hilbert space of quantum gravity. In particular, the highest entropy configurations, which account for almost all of the volume of semiclassical phase space, would not have quantum counterparts, i.e.~would not correspond to allowed states in a quantum theory of gravity.
\end{abstract}

\pacs{04.70.Dy, 04.40.--b, 04.20.Gz, 11.25.Tq}

\maketitle

{\bf I. Concepts}

\bigskip

All fundamental physical theories that we know, like quantum mechanics or general relativity, obey the principle of unitarity: knowledge about the state of a system at one instant is equivalent to knowledge about its state at any other instant. This is due to a one-to-one correspondence, induced by the evolution equations, between states at two different instants, and allows for prediction of the future and retrodiction of the past if the present state of the system is known. It is appealing to hypothesize that unitarity is itself a fundamental feature of Nature, especially in light of the AdS/CFT correspondence \cite{AdSCFTreview}, although it remains a challenge to understand \emph{how} it survives the unification of gravity and quantum mechanics, with the resulting phenomena of black hole evaporation \cite{HawkingPureMixed}.

Statistical (microcanonical) entropy $S$ is the logarithm of the number of distinct microstates $\psi$ of a system consistent with some imposed macroscopic properties. Thus, entropy $S$ is proportional to the dimensionality of the Hilbert space of allowed $\psi$'s and measures the amount of information that is encoded in a particular microstate $\psi$. Unitarity forbids any change in the size of this Hilbert space during evolution of the system, so, in particular, entropy cannot decrease (but may increase as we coarse-grain macroscopic information so that more microstates are accommodated). By this logic, the second law of thermodynamics is a consequence of unitarity.

Without a theory of quantum gravity, we do not know, so cannot count, the microstates of black holes (for results in string theory, see \cite{StringTheoryEntropy}). But it has been established semiclassically that a large black hole of mass $M$ emits thermal radiation of temperature $T \sim M^{-1}$ \cite{HawkingEntropy}, so the entropy in this Hawking radiation is of order the area of the hole: $S = \int dQ / T \sim \int dM\,M \sim A$ (we use Planck units $\hbar = c = G = 1$ throughout). Strictly speaking, the Hawking process applies only to the semiclassical part of the evaporation, but the final quantum part releases at most of order the Planck energy, which can be made negligible compared to the initial mass of the hole and is thus unlikely to change the scaling with $M$ of the total amount of radiation entropy \cite{ss}. A total black hole entropy of $S_{BH} = A/4$, corresponding to an entropy density $\sim 10^{69}\,{\rm bit}/{\rm m}^2$ on the horizon, is consistent with other evidence ranging from black hole thermodynamics \cite{BekensteinEntropy,HawkingEntropy} to string theory \cite{StringTheoryEntropy}, although there are other interpretations of this area entropy as well, e.g.~\cite{NatureOfBHEntropy}.

By unitarity or the second law, the entropy carried away in Hawking radiation cannot be smaller than the total black hole entropy if evaporation is \emph{complete} (as in Figure 1 \cite{fn}), i.e., if there are no remnants left at the end of the evaporation process and if no topology change occurs, whereby information that hit the singularity could escape into parts of the universe other than that containing the external observer \cite{TopologyChange}. The existence of Planck size remnants that contain significant entropy is aesthetically unpleasing since their number would have to be essentially infinite to account for the unradiated entropy; and in the AdS/CFT picture, topology change in a universe corresponds to an implausible scattering process in the field theory, in which large numbers of degrees of freedom somehow cease to interact (become causally disconnected).

\bigskip

{\bf II. The Puzzle}

\bigskip

In this essay we assume that physics, in particular physics describing black hole phenomena, is unitary and that black hole evaporation is complete. Lacking a quantum theory of gravity, these assumptions are unjustified, though perhaps well motivated.

Under these assumptions, all information (e.g., on the spacelike slice $\Sigma_0$ in Figure 1) which falls into a black hole has to re-emerge during evaporation in the Hawking radiation, thereby preserving unitary evolution without remnants or topology change. So there has to exist a one-to-one correspondence between the future states of the universe on future infinity $\mathscr{I}^+ \cup i^+$ and the past states of the universe on any earlier Cauchy slice (e.g., one, like $\Sigma_0$, prior to the black hole). This requires the future and past Hilbert spaces to have the same dimensionality.

\begin{figure}[t]
\includegraphics[height=10.3cm]{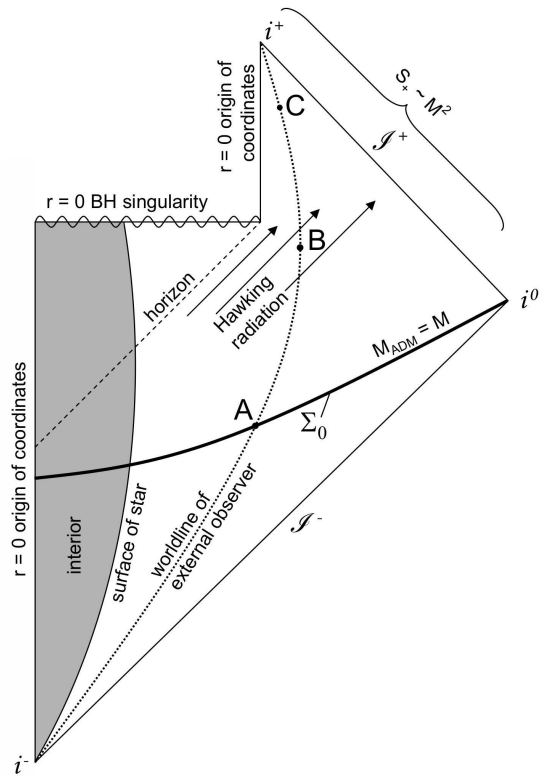}
\caption{Penrose diagram of a collapsing spherically symmetric object of mass $M$: as it contracts, a horizon forms behind which eventually all matter falls into the black hole singularity $r=0$. Initially (event A), an outside observer sees a shrinking star; after a while (B), her observations are consistent with the existence of a black hole, i.e.~dim red-shifted light from the frozen surface and outgoing thermal Hawking radiation, which causes decrease in the central mass; finally (C), all Hawking radiation has passed the observer and the black hole has evaporated completely, leaving behind empty space with all information in the radiation at $\mathscr{I}^+$ and $i^+$. Also shown is a spacelike Cauchy slice $\Sigma_0$ preceding the black hole. Under our assumptions of unitarity and complete evaporation there exists a one-to-one correspondence between states $\Sigma_0$ (matter+gravity) and states on future infinity $\mathscr{I}^+ \cup i^+$.}
\end{figure}

And it allows us to answer the question: how big is the Hilbert space of states that can form a black hole of mass $M$? It has the same dimension as the future Hilbert space on $\mathscr{I}^+ \cup i^+$, i.e.~the same as the space of Hawking radiation states. Thus, the entropy $\sim M^2$ in the black hole radiation implies a bound on the entropy of the black hole precursor, or equivalently the dimension of its Hilbert space (we assume an initially pure state).

But: there exist semiclassical configurations $\Sigma_0$ (matter+gravity) that collapse to form a black hole of mass $M$ and have entropy $S_{\Sigma_0}$ much larger than $\sim M^2$ (i.e., much larger than that of a black hole of equal ADM mass). We will describe examples below.

The assumptions of unitarity and complete evaporation therefore force the exclusion of highly entropic ($S > M^2$) semiclassically allowed states from the Hilbert space of quantum gravity. The, perhaps surprising, result is that not all semiclassical configurations (at fixed ADM mass) have quantum counterparts, i.e.~the vast majority do not correspond to allowed states in a quantum theory of gravity.

Our argument is logically distinct from the black hole information problem as usually considered (evolution of an initially pure state into a mixed state after evaporation \cite{HawkingPureMixed}; see also \cite{susskindlindesay}) -- we \emph{assume} unitary evolution and do not attempt to address \emph{how} the information manages to escape the black hole. Rather, our result is a statement about the Hilbert space of quantum gravity based on fundamental considerations. And although it can be cast in the form of an entropy bound -- restricting the amount of entropy allowed in a black hole precursor -- the specific high entropy configurations which are excluded do \emph{not} violate Bousso's covariant entropy bound \cite{boussoreview,boussobound}.

\bigskip

{\bf III. High Entropy Configurations}

\bigskip

Now we present two examples of classes of configurations $\Sigma_0$ (matter+gravity) that cause the puzzle outlined above. In both examples, the curvature of space on $\Sigma_0$ makes the ADM mass of the configuration (i.e., the energy a distant external observer sees and that determines the black hole area and hence Hawking radiation entropy) much smaller than would be suspected from the proper internal volume, to which the initial entropy $S_{\Sigma_0}$ is proportional. In the case of example (a) (``monsters''), this effect can be ascribed to large negative binding energy \cite{Hsuzero} which almost cancels the proper mass to yield a relatively small ADM mass. In (b), the Kruskal-FRW example, the reason is the non-monotonic behavior of the radius $r$ of 2-spheres across the outer Einstein-Rosen bridge.

\bigskip

{\bf (a) Monsters}

\bigskip

Our first example is a ball of material which is on the verge of collapsing to form a black hole. Its energy density profile is arranged to produce a curved internal space with large proper volume (see Figure 2(a)). The configuration is spherically symmetric, defined by initial data on a Cauchy slice $\Sigma_0$ at a moment of time symmetry (i.e., configuration initially ``at rest'') without (marginally) trapped surfaces, so that $\Sigma_0$ has geometry
\begin{equation}
\label{MetricSigma0}
ds^2 \bigr\vert_{\Sigma_0}
= \epsilon(r)^{-1} dr^2+r^2 d \Omega^2~,\quad K_{ab} \bigr\vert_{\Sigma_0} = 0~,
\end{equation}
with $\epsilon(r) > 0$. For given initial matter distribution $\rho(r)$, Einstein's (constraint) equations determine (e.g., \cite{WaldGRbook})
\begin{equation}
\epsilon(r)=1-\frac{2M(r)}{r}~,
\end{equation}
where
\begin{equation}
M(r) = 4 \pi \int_0^r dr' \, {r'\,}^2 \rho(r')~.
\end{equation}
If a configuration has radius $R$, i.e.~$\rho(r>R) = 0$, its ADM energy is $M = M(R)$. This quantity is constant during time evolution of the configuration (Birkhoff's theorem), and, if it collapses to a black hole, equals its mass.

\begin{figure}[t]
\includegraphics[width=14cm]{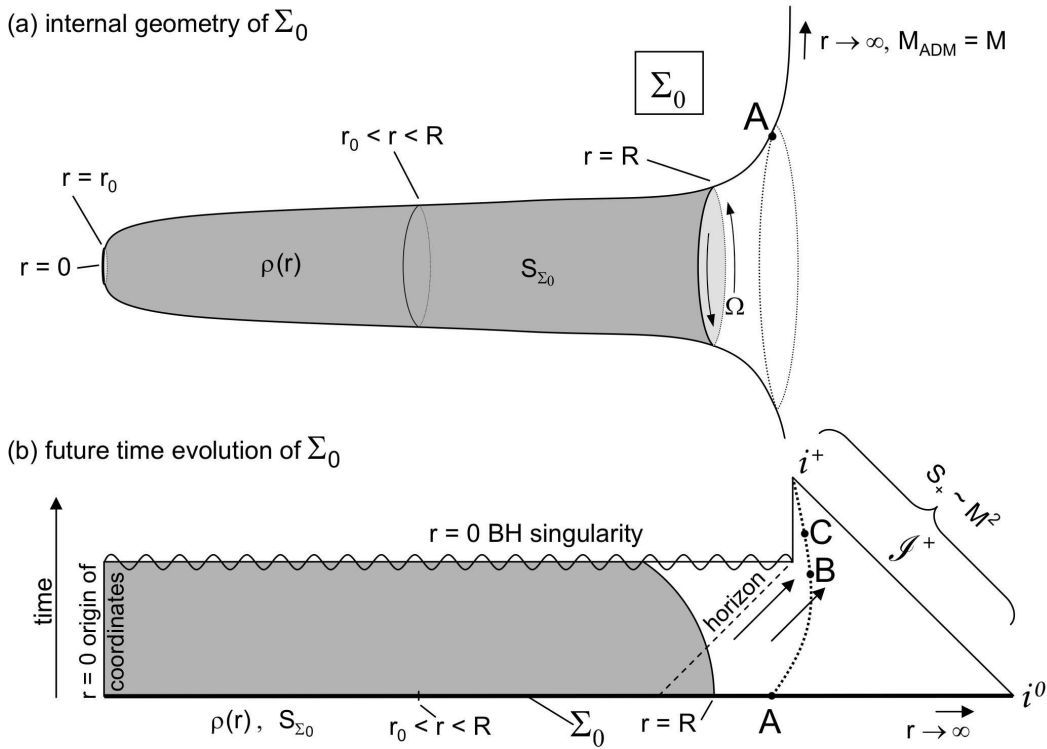}
\caption{(a) Embedding of the monster configuration $\Sigma_0$ into flat space with one angular dimension suppressed. The ``neck'' has proper length much bigger than $\left(R-r_0\right)$, due to the huge factor $\epsilon(r)^{-1/2}$, and contains all of the initial entropy $S_{\Sigma_0}$. For $r>R$ the geometry is just that of a Schwarzschild slice with mass $M = M_{ADM}$. (b) The monster's future time evolution is similar to ordinary gravitational collapse (cf. Figure 1): (almost) all matter and entropy, if it was not already initially, will fall behind a horizon (infall of outer layers soon creates trapped surfaces) and form a black hole which then evaporates, radiating away entropy $S_+ \sim M^2 < S_{\Sigma_0}$ past the external observer to future infinity $\mathscr{I}^+ \cup i^+$.}
\end{figure}

Now, consider a semiclassical configuration (``\emph{monster}'' \cite{monsterpaper,SorkinWaldZhang}, Figure 2(a)) with radius $R \gg 1$ that yields
\begin{equation}
\epsilon(r)=\left(\frac{r_0}{r}\right)^{\gamma}~,\quad\quad r_0 < r < R~,
\end{equation}
with some $\gamma>0$ and $r_0 \ll R$ (to avoid poles), so that the configuration comes increasingly closer to forming trapped surfaces as $r \nearrow R$ (long ``neck'' in Figure 2(a)). It has ADM mass
\begin{equation}
M = \frac{R}{2}\left(1-\epsilon(R)\right) \approx \frac{R}{2} \sim R
\end{equation}
and energy density
\begin{equation}
\rho(r)=\frac{M'(r)}{4\pi r^2}
\approx \frac{1}{8\pi r^2} \sim \frac{1}{r^2}~,\quad\quad r_0 < r < R~.
\end{equation}
Finally, with a relation $s = \alpha \rho^{\beta} \sim \rho^{\beta}$ between energy and entropy density of the matter ($\alpha = {\cal O}(1)$), the initial entropy is
\begin{equation}
S_{\Sigma_0} = 4 \pi \int_0^R dr \, r^2 \epsilon(r)^{-1/2} s(r)  \sim  \frac{R^{3-2\beta+\gamma/2}}{r_0^{\,\gamma/2}} \sim A^{3/2-\beta+\gamma/4}~,
\end{equation}
with $A \sim M^2$ the area of the black hole formed in collapse of this monster.

It is now evident that, if $\beta$ is constant, one can always find configuration parameters $\gamma$ such that the entropy of the monster exceeds area scaling (hence, the name). This is the case, e.g., if we model the matter (initially) as a perfect fluid with equation-of-state parameter $w$. Then $\beta = 1/\left(1+w\right)$, and we would just have to choose $\gamma > 1$ for a photon gas ($w = 1/3$) or $\gamma > 2$ for dust ($w = 0$; we assume the dust particles carry some kind of label or have spin).

Figure 2(b) depicts the time evolution of a monster, which resembles ordinary gravitational collapse (Figure 1). The main difference is that, due to our construction, the entropy $S_{\Sigma_0}$ on the initial Cauchy slice can be much bigger than the entropy $S_+$ on future infinity. This monster is therefore a semiclassical configuration with \emph{no} corresponding microstates in a quantum theory of gravity. Note, if $r_0$ is chosen a few orders of magnitude above the Planck length, all involved densities $\rho(r)$ and $s(r)$ are sub-Planckian, so that our semiclassical analysis naively applies. Furthermore, Bousso's covariant entropy bound \cite{boussobound} holds in the semiclassical monster spacetime since it falls under the general class of spacetimes for which the theorem of Flanagan, Marolf and Wald \cite{FMW} applies. (This assumes no large entropy gradients due to, e.g., shockwaves during evolution, which seems plausible, but has not been proven.)

\bigskip

{\bf (b) Kruskal-FRW gluing}

\bigskip

The second example consists of slices of closed FRW universes which are glued together across Einstein-Rosen bridges, eventually connecting to a large asymptotically flat universe (Figure 3(a)). Again, a larger proper volume can be accommodated at fixed ADM mass. The configuration is specified, as before, by initial data on a spherically symmetric and time symmetric ($K_{ab} \vert_{\Sigma_0} = 0$) Cauchy slice $\Sigma_0$: we take the part of a constant-time slice of the Kruskal spacetime with mass $M_1$ (e.g., part of the $U+V = 0$ slice, in usual Kruskal coordinates) that contains one asymptotic region with outside observer A, the Einstein-Rosen bridge at its maximal extent $r=2M_1$ and the piece $r_{1{\rm l}} > r > 2M_1$ of the other asymptotic region (right part in Figure 3(a)). This is then glued onto the part $\chi < \chi_{1{\rm l}}$ of the hypersurface $ds^2 = a_{12}^{\,2} \left( d\chi^2+\sin^2 \chi \, d\Omega^2 \right)$ representing a closed FRW universe at the instant of its maximal expansion $a_{12}$. By cutting this 3-sphere off at $\chi = \chi_{2{\rm r}}$, a second piece of Kruskal containing an Einstein-Rosen bridge can be joined, etc. In our notation the integer subscript $n$ denotes the $n$-th Einstein-Rosen bridge, and l,r denote left,right (see Figure 3). 

Matching the geometry across the common boundary requires the transverse metric to be continuous and continuously differentiable (i.e.~the extrinsic curvature $K^{(3)}_{ab}$ has to be the same on either side); its second derivative can be discontinuous, as is the energy density $\rho$, consistent with Einstein's equation $G_{ab} = 8\pi T_{ab}$. At the rightmost joining surface in Figure 3(a), continuity of the transverse metric means equality of the areas of the spherical sections $\chi=\chi_{1{\rm l}}$ and $r=r_{1{\rm l}}$, i.e.
\begin{equation}
\label{equalA}
a_{12}\sin \chi_{1{\rm l}} = r_{1{\rm l}}~.
\end{equation}
And equality of extrinsic curvatures is, in the case of spherical symmetry, equivalent to continuous differentiability of the area $A(R)$ of 2-spheres with respect to proper radial distance $R$:
\begin{equation}
\frac{d}{a_{12}\,d\chi} \left( 4\pi a_{12}^{\,2} \sin^2 \chi \right) \biggr\vert_{\chi=\chi_{1{\rm l}}} \,=\, 
\frac{d}{- \left( 1-2M_1/r \right)^{-1/2}\,dr} \left( 4\pi r^2 \right) \biggr\vert_{r=r_{1{\rm l}}}~,
\end{equation}
which forces $\chi_{1{\rm l}} \in [\pi/2,\pi)$ and, with (\ref{equalA}),
\begin{equation}
\label{diffA}
2M_1 = a_{12}\sin^3\chi_{1{\rm l}}~.
\end{equation}
Equations like (\ref{equalA}) and (\ref{diffA}) hold at every joining surface, with a modified constraint $\chi_{{\rm r}} \in [0,\pi/2]$ if joining just \emph{right} of an Einstein-Rosen bridge. From these formulae, a configuration like Figure 3(a) can be constructed, e.g., in the following way: first pick masses $M_1,~M_2, \ldots$ describing the Kruskal pieces ($M=0$ forces the construction to an end), then sizes $a_{12},~a_{23} \ldots$ of the FRW pieces subject to constraints $a_{12} \ge 2M_1,~2M_2$, etc. $\Sigma_0$ is then uniquely determined.

\begin{figure}[t]
\includegraphics[width=14cm]{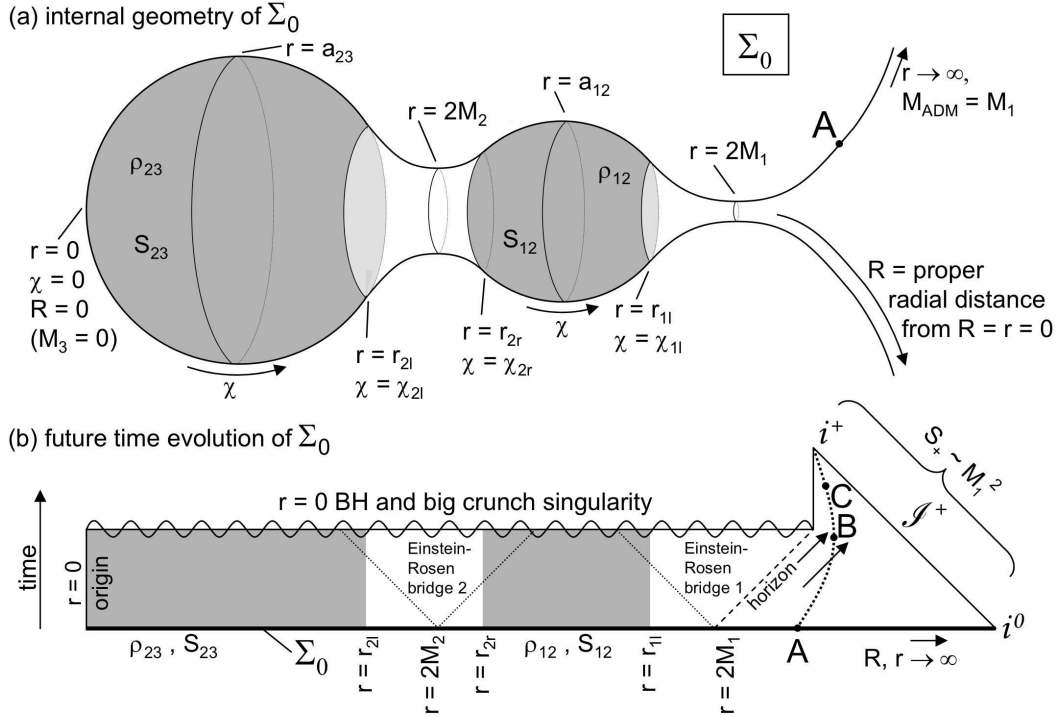}
\caption{(a) Embedding of glued Kruskal-FRW initial slice $\Sigma_0$ into flat space with one angular dimension suppressed. $R$ is the proper radial distance from the innermost point and $r=r(R)$ gives the radius of the 2-sphere labeled $R$. Additional or larger closed FRW pieces could be adjoined, and there could also be a second asymptotic Kruskal piece (even with mass parameter different from $M_1$) if the far left were not closed off with a 3-sphere. (b) By considering the rightmost Einstein-Rosen bridge, standard energy conditions suffice to show that a singularity will form and that the external observer will see a black hole of mass $M_1$ whose Hawking radiation then contains potentially much less entropy than was present on $\Sigma_0$. In the case of pressureless dust, the time evolved spacetime can be given analytically as Kruskal spacetimes and FRW universes appropriately sewn together (Oppenheimer-Snyder collapse \cite{RelativistsToolkit,MTW}).}
\end{figure}

Invoking Friedmann's equation with vanishing instantaneous expansion, the FRW pieces have energy density $\rho_{12}=3/8\pi a_{12}^{\,2}$. With $s \sim \rho^{\beta}$, the entropy of one piece becomes
\begin{equation}
\label{S12equation}
S_{12} = 4 \pi a_{12}^{\,3}s \int_{\chi_{2{\rm r}}}^{\chi_{1{\rm l}}} d\chi \, \sin^2\chi \, \sim\,
a_{12}^{\,3-2\beta} \left[ \chi_{1{\rm l}}-\chi_{2{\rm r}}-\frac{1}{2}\sin 2\chi_{1{\rm l}} +\frac{1}{2}\sin 2\chi_{2{\rm r}} \right].
\end{equation}
The bracket in (\ref{S12equation}) approaches $\pi = {\cal O}\left(1\right)$ as $a_{12}$ becomes a few times bigger than $2M_1$ and $2M_2$. In that case, the total entropy on $\Sigma_0$ is
\begin{equation}
\label{initialSFRW}
S_{\Sigma_0} =
S_{12}+S_{23}+\ldots \,\sim\,
a_{12}^{\,3-2\beta} + a_{23}^{\,3-2\beta}+\ldots~~,
\end{equation}
and so can be made arbitrarily big (for any $\beta=1/\left(1+w\right)<3/2$) by either taking the size of the FRW pieces or their number to be large.

Evolved forward in time (Figure 3(b)), the entropy in the Hawking radiation that passes the external observer and reaches future infinity is $S_+ \sim M_1^{\,2}$, so again is potentially much less than the entropy on the initial slice (\ref{initialSFRW}). As in the case of monsters, the Kruskal-FRW configurations are reasonable semiclassical initial data insofar as all involved densities are well sub-Planckian (if the FRW pieces are a few orders of magnitude bigger than the Planck length).  The spacetimes do not violate the covariant entropy bound by the same arguments \cite{FMW} as before (see also \cite{boussobound} for more specific discussion of entropy bounds in closed FRW universes).

\bigskip

Both types of configurations have the pathological property that, under isolated evolution, they must have emerged from a past singularity (white hole). This can be seen via backward evolution of the time symmetric initial data, noting that forward evolution leads to a black hole and future singularity. Furthermore, relaxing the assumption of isolation, the configurations cannot be constructed ``in the laboratory,'' even via intervention by an arbitrarily advanced civilization \cite{monsterpaper,SorkinWaldZhang}. Despite their pathologies, these configurations represent valid semiclassical states of a matter-gravity system: they are all locally well behaved, in particular do not require large energy or entropy densities, and (if present in the Hilbert space) could be accessible via tunneling starting from an ordinary matter configuration with the same quantum numbers (ADM energy, angular momentum, charge). 

\bigskip

{\bf IV. Conclusion}

\bigskip

Under the assumptions that physics is unitary and that black hole evaporation is complete, we have shown that the Hilbert space of quantum gravity is much smaller than expected based on semiclassical considerations. The overwhelming majority (as defined by phase space volume or entropy) of semiclassical matter-gravity configurations are not allowed states in the corresponding quantum theory of gravity.

Presumably some new physical principle is required to exclude these high entropy states. This might be due to the very formulation of quantum gravity, e.g., if it is formulated as a lower-dimensional theory on future infinity, due to holography. Note that in both of our high entropy examples one needs several \emph{disconnected} holographic screens \cite{holographicscreens} on which to project all of the spacetime. Whatever the new principle is, it must be global in nature: the excluded configurations have no exceptional characteristics when considered locally.

\bigskip

\emph{Acknowledgments ---} The authors thank Sean Carroll, Nick Evans, Ted Jacobson, Hirosi Ooguri, John Preskill and Mark Wise for useful discussions. The authors are supported by the Department of Energy under DE-FG02-96ER40969.


\begin{thebibliography}{99}

\bibitem{AdSCFTreview}
For a review, see
O.~Aharony, S.~S.~Gubser, J.~M.~Maldacena, H.~Ooguri and Y.~Oz,
``Large N field theories, string theory and gravity,''
Phys.~Rept.~{\bf 323}, 183 (2000)
[hep-th/9905111].

\bibitem{HawkingPureMixed}
S.~W.~Hawking,
``Breakdown of predictability in gravitational collapse,''
Phys.~Rev.~D {\bf 14}, 2460 (1976).

\bibitem{StringTheoryEntropy}
A.~Strominger and C.~Vafa,
``Microscopic origin of the Bekenstein-Hawking entropy,''
Phys.~Lett.~B {\bf379}, 99 (1996)
[hep-th/9601029];
J.~M.~Maldacena, A.~Strominger and E.~Witten,
``Black hole entropy in M theory,''
JHEP {\bf 12}, 002 (1997)
[hep-th/9711053].


\bibitem{HawkingEntropy}
S.~W.~Hawking,
``Particle creation by black holes,''
Commun.~Math.~Phys.~{\bf 43}, 199 (1975).

\bibitem{ss}
For a discussion of possible corrections to the semiclassical entropy result, see, e.g., 
S.~Shankaranarayanan,
 ``Do subleading corrections to Bekenstein-Hawking entropy hold the key to quantum gravity?,''
  arXiv:0805.4531 [gr-qc].


\bibitem{BekensteinEntropy}
J.~D.~Bekenstein,
``Black holes and entropy,''
Phys.~Rev.~D {\bf 7}, 2333 (1973).

\bibitem{NatureOfBHEntropy}
T.~Jacobson,
``On the nature of black hole entropy,''
gr-qc/9908031.

\bibitem{fn} In our discussion we assume a background spacetime which is large enough to allow, e.g., total evaporation of a large, semiclassical black hole. Thus, we do not address the case of the most general semiclassical spacetimes, which may not be asymptotically flat or large.

\bibitem{TopologyChange}
S.~D.~H.~Hsu,
``Spacetime topology change and black hole information,''
Phys.~Lett.~B {\bf 644}, 67 (2007)
[hep-th/0608175].

\bibitem{susskindlindesay}
L.~Susskind and J.~Lindesay, An introduction to black holes, information and the string theory revolution: The holographic universe, {\it World Scientific, Hackensack, USA (2005)}.

\bibitem{boussobound}
R.~Bousso,
``A covariant entropy conjecture,''
JHEP {\bf 07}, 004 (1999)
[hep-th/9905177].

\bibitem{boussoreview}
R.~Bousso,
``The holographic principle,''
Rev.~Mod.~Phys.~{\bf 74}, 825 (2002)
[hep-th/0203101].

\bibitem{Hsuzero}
S.~D.~H.~Hsu,
``Zero energy configurations in general relativity,''
gr-qc/9801106.

\bibitem{WaldGRbook}
R.~M.~Wald, General Relativity, {\it The University of Chicago Press, Chicago, USA (1984)}, chapters 6.2 and 10.2.

\bibitem{monsterpaper}
S.~D.~H.~Hsu and D.~Reeb,
``Black hole entropy, curved space and monsters,''
Phys.~Lett.~B {\bf 658}, 244 (2008)
[arXiv:0706.3239 [hep-th]].

\bibitem{SorkinWaldZhang}
R.~D.~Sorkin, R.~M.~Wald and Z.~J.~Zhang,
``Entropy of self-gravitating radiation,''
Gen.~Rel.~Grav.~{\bf 13}, 1127 (1981).

\bibitem{FMW}
\'E.~\'E.~Flanagan, D.~Marolf and R.~M.~Wald,
``Proof of classical versions of the Bousso entropy bound and of the generalized second law,''
Phys.~Rev.~D {\bf 62}, 084035 (2000)
[hep-th/9908070].

\bibitem{RelativistsToolkit}
E.~Poisson, A Relativist's Toolkit -- The Mathematics of Black-Hole Mechanics, {\it Cambridge University Press, Cambridge, UK (2004)}, chapter 3.8.

\bibitem{MTW}
C.~W.~Misner, K.~S.~Thorne and J.~A.~Wheeler, Gravitation, {\it W.~H.~Freeman and Company, San Francisco (1973)}, chapter 32.4.

\bibitem{holographicscreens}
R.~Bousso,
``Holography in general space-times,''
JHEP {\bf 06}, 028 (1999)
[hep-th/9906022].

\end{thebibliography}
\end{document}